%Paper: cond-mat/9503171
%From: Alon Drory <drory@buphyk.bu.edu>
%Date: Fri, 31 Mar 1995 16:05:03 -0500

\overfullrule=0 pt
\magnification 1200
\baselineskip=18 true bp

%----------------definitions------------------

\def\eg{{\it e.\ g.}}\def\ie{{\it i.\ e.}}\def\apr{{\it a priori}}
\def\ref#1{${}^{#1}$}
\newcount\refnum\refnum=0  %automatic, non-labelled references
\def\refi{\smallskip\global\advance\refnum by 1\item{[\the\refnum]}}

\def\pd#1#2{{\partial #1\over\partial #2}}      %partial derivative
\def\p2d#1#2{{\partial^2 #1\over\partial #2^2}} %second partial derivative
\def\td#1#2{{d #1\over d #2}}      %total derivative
\def\t2d#1#2{{d^2 #1\over d #2^2}} %second total derivative
\def\av#1{\langle #1\rangle}                    %average

\def\jsp #1 #2 #3 {{\sl J. Stat.\ Phys.} {\bf #1}, #2 (#3)}
\def\pra #1 #2 #3 {{\sl Phys.\ Rev.\ A} {\bf #1}, #2 (#3)}
\def\pre #1 #2 #3 {{\sl Phys.\ Rev.\ E} {\bf #1}, #2 (#3)}
\def\prl #1 #2 #3 {{\sl Phys.\ Rev.\ Lett.} {\bf #1}, #2 (#3)}
\def\rmp #1 #2 #3 {{\sl Rev.\ Mod.\ Phys.} {\bf #1}, #2 (#3)}

\def\dt{\Delta t}
\def\d{\delta}
\def\k{\mu}
\def\p{\phi}
\def\f{\varphi}
\def\dx{\d (x-x_n)}
\def\dv{\d (v-v_n)}
\def \di#1{\partial_{1} #1}
\def\Wnt{W_n(y_1,t_1;\, y_2,t_2;\ldots;\, y_n,t_n)}
\def\Wni{W_n(y_1,\dt ;\, y_2,2\dt ;\ldots;\, y_n, n\dt)}
\def\Wn{W_n(y_1,\ldots, y_n)}
\def\gtt{g(t,t')}
\def\htt{h(t,t')}
\def\th{\theta}
\def\fpn{\hat p(k ,\th ,n)}
\def\fpt{\hat p(k ,\th ,t)}
\def\wn{\widehat W(\f_1 ,\ldots,\f_n)}
\def\yy{y_{i+1}-\k (\dt)y_i}
\def\po{P_0(y_1)}
\def\Int{\int^{\infty}_{-\infty}}
\def\li {n \to \infty, \dt \to 0}
\def \h {{1 \over 2}}
\def \pdd#1#2{{\partial^2 #1 \over \partial #2 ^2}}
%---------------------------------------------

\centerline{\bf Theory of Second and Higher Order Stochastic Processes}
\bigskip
\centerline{\bf Alon Drory}
\smallskip
\centerline{Center for Polymer Studies and Department of Physics}
\centerline{Boston University, Boston, MA 02215}
\vskip 1in
\centerline{ABSTRACT}
{\narrower\narrower\smallskip\noindent}
This paper presents a general approach to linear stochastic processes
driven by various random noises. Mathematically, such processes are
described by linear stochastic differential equations of arbitrary
order (the simplest non-trivial example is $\ddot x = R(t)$, where
$R(t)$ is not a Gaussian white noise). The stochastic process is
discretized into $n$ time-steps, all possible realizations
are summed up and the continuum limit is taken. This procedure often
yields closed form formulas for the joint probability
distributions. Completely worked out  examples include  all Gaussian
random forces and a large class of Markovian (non-Gaussian)
forces. This approach is also useful for deriving Fokker-Planck
equations for the probability distribution functions. This is worked out
for Gaussian noises and for the Markovian dichotomous noise.

{\narrower\bigskip\noindent
P. A. C. S. Numbers: 05.40.+j., 02.50.-r.}

\vfill\eject
\medskip\centerline{\bf I. Introduction}\smallskip

Stochastic differential equations were first studied in the context of
Brownian motion by Langevin. For a particle of unit mass, he derived
the equation for the displacement
$$
\ddot x + \gamma \dot x = R(t)\, .\eqno(1.1)
$$
In this equation, the drag force, $-\gamma \dot x $, is a deterministic
term which represents the ``environment'' in which the particle moves.
In the case of Brownian motion, this is the fluid surrounding the
particle. The force driving the particle through this environment is the
random term $R(t)$ which is assumed to be a Gaussian white noise [1].

In recent years, the field of stochastic equations has expanded
considerably along two main lines, each focusing on one of the two basic
aspects of Langevin equations: The first concentrates on more complex
environments, usually introduced through potentials $U(x)$; the second
concentrates on random driving forces more complex than the Gaussian
white noise.

The first approach deals with equations like, \eg,
$$
\ddot x +\td {U(x)} x  =R(t)\, ,\eqno(1.2)
$$
where $U(x)$ is an external potential [2]. Since Eq.~(1.2) is typically
non-linear and very hard to solve, the random noise is usually kept
Gaussian and white. (As an aside, note that when one tries to combine
non-trivial potential terms with colored noises [3,4,5], the
difficulties are so great that the order of the equation has to be
reduced to the form [4] $\dot x + \Omega (x) =R(t)\, $).

The present work, however, follows the second line of generalizations of
Eq.~(1.1), which leaves the ``environment'' relatively simple but
considers more complex random forces without reducing the order of the
equation. It has been known for some time, for example,
that a more realistic description of Brownian motion must include
a finite correlation time for the random force [6], a feature absent in
the Gaussian white noise. If the random force arises from processes
internal to the system (as in Brownian motion), the system may reach a
state of detailed balance at equilibrium. In this case, if the force is
not Gaussian white, the friction term requires a modification which
usually involves a retardation effect [7]. Eq.~(1.1) is then generalized
to be :
$$
\ddot x +\int_{-\infty}^t \!\!\gamma (t-t') \dot x (t')\, dt' =
R(t)\, ,\eqno(1.3)
$$
where $\gamma(t-t')$ is determined by the correlation function
$\av {R(t)R(t')}$ (hereafter, angular brackets denote average
values) [7]. Unlike the
generalization Eq.~(1.2), Eq.~(1.3) is linear in the stochastic
variable $x(t)$. Nonetheless, it is usually still difficult to solve.
Often, it is simplified by assuming the friction to be negligible (this
cannot be done if detailed balance is required. However, in many
systems, this requirement is not necessary because the force is an
external influence). One then ends up with the simple equation
$$
\ddot x =R(t)\,  .\eqno(1.4)
$$
This approximation has been used recently to describe some
reaction-diffusion processes in a force-dominated regime [8].

Even the simple equation (1.4) is still hard to solve when the random
force is more complex than Gaussian white noise. In this context,
solving the equation means finding the joint probability distribution
function (pdf) for the position and the velocity, $p(x,v,t)$,
the marginal distribution for the velocity, $p(v,t) \equiv \int p(x,v,t)
dx$, and the marginal distribution for the position, $p(x,t) \equiv \int
p(x,v,t) dv$. Eq.~(1.4) does not relate to these functions directly,
however, and the usual approach is therefore to obtain a Fokker-Planck
(FP) equation for the pdf's. Going from a stochastic equation like
(1.4) to a FP equation can be quite difficult. After this is done, one
still has to solve the resulting FP equation, which is far from trivial
[9,10,11].

In the present paper, I outline a general approach to linear stochastic
equations (including Eq.~(1.3) and (1.4)), that often yields a closed
form expression for the various pdf's, and may also simplify the
derivation of Fokker-Planck equations.

This approach is based on a {\it discretization} of the stochastic
process represented by the Langevin equation by breaking up the
continuous time variable into a finite number of steps, $n$. The
discrete process thus generated is completely described by an $n$-order
distribution function $W_n$. In section II, I derive the fundamental
formula of this paper, which yields p(x,v,t) as the limit $n \to
\infty$ of an expression involving $W_n$. I do not attempt a
rigorous mathematical proof that the limit exists in general. This would
be an extremely difficult problem since the expression obtained in
section II is basically a path integral. However, the path integral
formalism is entirely bypassed in this work in favor of a direct
calculation approach. Indeed, the remaining sections of the paper are
devoted to specific examples where the calculations can be carried
through without using any path-integral method.

Thus, in section III, the fundamental formula is applied to a class of
Markov noises termed {\it monovariant}. This includes the
Ornstein-Uhlenbeck noise, as well as, \eg, the Wiener and Cauchy
processes. In section IV, the same formula yields a closed expression
for the various pdf's when the noise is any Gaussian process (not
necessarily Markovian). Finally, I use the general formula to derive
FP-like equations: Section V treats Gaussian forces, and section VI
treats the Markovian dichotomous  noise, with which many investigations
were concerned in recent years. Section VII summarizes the approach and
its main results.

\bigskip\centerline{\bf II. General Formalism}\smallskip

Eqs.~(1.3) and (1.4) are particular cases of the general equation
$$
L(x,t)=R(t)\, ,\eqno(2.1)
$$
where $L(x,t)$ is an operator acting on the stochastic variable $x$. I
assume that $L(x,t)$ is such that the general solution of
the equation can be written formally as
$$
x=u(t) + \int_0^t \gtt R(t') dt'\, .\eqno(2.2)
$$
$u(t)$ is the solution to the homogeneous equation, and $\gtt$ is
the Green's function of the equation. Often, $\gtt$ depends
only on the combination $t-t'$, but this is not yet assumed.

Clearly, a similar relation holds for the velocity as well. Thus, we can
find functions $w(t)$ and $\htt$ such that
$$
v(t) = w(t) + \int_0^t \htt R(t') dt' \, ,\eqno(2.3)
$$

Eqs.~(2.2) and (2.3), rather than Eq.~(2.1), are the starting point of
our approach. Thus, in the remainder of this paper, I assume that $u(t),
w(t), \htt$ and $\gtt$ are given functions (though they may be hard to
determine in practice).  For simplicity, I also assume throughout the
paper that $\av {R(t)} =0$. The modifications required to accommodate a
non-zero mean value are straightforward but tedious.

The number of stochastic variables of interest is usually determined by
the order of the stochastic equation. Thus, in a second order process
such as described by $\ddot x = R(t)$, we have two variables of
interest, $x(t)$ and $v(t)$. However, the method described here is
general and applicable to any number of stochastic variables and
therefore to equations of arbitrary order. However, for definiteness, I
will describe it in the context of second order processes, which are the
most common in physics.

The main idea presented here is to transform the continuous process
described by Eqs.~(2.2)-(2.3) into a discrete process involving finite
time steps $\dt$. Such discrete processes can be described easily
through the $n$-order distribution function of the random noise $R(t)$.
It turns out that the Fourier transform of this distribution function
relates very simply to the Fourier transform of the pdf of the
stochastic variable. This basic relation can be used to calculate the
pdf explicitly in some cases, or to easily derive Fokker-Planck-like
equations in other cases.

To derive this fundamental relation, note that the discretized version
of the random noise $R(t)$ is fully determined by the set of all its
$n$-order distribution functions
$$
\Wnt \,dy_1\cdots dy_n \equiv {\rm Prob}\ \{
y_i< R(t_i) < y_i+dy_i\enspace {\rm for\ all}\enspace i=1\cdots
n\}\,.\eqno(2.4)
$$
\ie , $\Wnt\,dy_1\cdots dy_n$ is the probability that at the
times $t_i$, the value of the random noise $R(t_i)$ is in the range
$(y_i, y_i+dy_i)$. The functions $W_n$ define the
full process $R(t)$ through discrete and finite sets of times $\{
t_i\}_{i=1}^n$. This suggests describing the random variables $x(t),
v(t)$ through similar finite sets of times. To this end, define a time
step $\dt=t/n$, where $n$ is an integer that will eventually go to
infinity.

Consider now a {\it particular} realization of the process, \ie, a
specific set of force values $\{y_i=R(i\dt)\}_{i=1}^n$. For this
realization, define sets $\{x_i\}_{i=1}^n$ and $\{v_i\}_{i=1}^n$ through
a discretization of Eqs.~(2.2) and (2.3),
\ie,:
$$
\eqalignno{
x_n&=u_n + \sum_{j=1}^n g(n,j)\,y_j\dt\, , &(2.5a)\cr
v_n &= w_n + \sum_{j=1}^n  h(n,j)\,y_j\dt \, , &(2.5b)\cr
\noalign{\hbox {where}}
u_n& = u(n\dt )\, , \qquad w_n = w(n\dt)\, , &(2.5c)\cr
g(n,j)& = g(n\dt , j\dt )\, , \qquad h(n,j) = h(n\dt , j\dt )\, .
&(2.5d)\cr
}
$$
As $n \to \infty$ and $\dt \to 0$ with $n\dt$ remaining constant, the
sets $\{x_i\}$ and $\{v_i\}$ converge to the continuous processes
$x(t),\, v(t)$.

Let us now define a discrete analog of the joint pdf  $\>p(x,v,t)$. This
discrete pdf, denoted $\>p(x,v,n)$, is defined as
$$
p(x,v,n) \,dx\, dv = {\rm Prob}\ \{ x < x_n < x + dx \quad {\rm
and}\quad v<v_n<v+dv \}\, ,\eqno(2.6)
$$
where $x_n$ and $v_n$ are now {\it defined} by Eqs.~(2.4) and (2.6),
with respect to the values ${y_i}$ of the random noise $R(t)$. Our aim
is to calculate $p(x,v,n)$ explicitly and hope that as $\li$, $p(x,v,n)$
will tend to $p(x,v,t)$.

The probability $p(x,v,n)$ is the total probability of all possible
realizations $\{y_i\}$ such that $x<x_n<x+dx$ and $v<v_n<v+dv$, where
$x_n$ and $v_n$ are given by Eqs.~(2.4)-(2.6) and $n=t/\dt$. For
simplicity, I assume that the set $\{y_i\}_{i=1}^n$ may take any value
in the range $(-\infty,\infty)$ (if the range is discrete, one can use
appropriate $\d$-functions to restrict the values of $y_i$). Therefore,
we have
$$
p(x,v,n) = \Int \! dy_1\cdots \!\Int \! dy_n \, \Wni \dx \dv\,
.\eqno(2.7)
$$
The two $\d$-functions in the integral, $\dx$ and $\dv$,
ensure that only processes $\{y_i\}_{i=1}^n$ which yield the appropriate
final positions and velocities are counted. Note that $x_n$ and $y_n$
are functions of $\{y_i\}$ through the relations (2.5). For simplicity, I
shall write from now on $\Wn$ instead of $\Wni$.

Eq.~(2.7) can be simplified by introducing the Fourier transforms (FT),
$$
\eqalignno{
\fpn& = \Int \!\! dx \Int \!\! dv \> e^{i(xk +v\th )}\,p(x,v,n)
\, .&(2.8a)\cr
\wn &= \Int\!\! dy_1 \cdots\! \Int\!\! dy_n \> \exp\left[i(y_1\f_1
+\cdots +y_n\f_n)\right]\,\Wn \, . &(2.8b)\cr
}
$$

Substituting Eq.~(2.7) into Eq.~(2.8) and interchanging the order of
integration, we obtain:
$$
\fpn = \Int\!\! dy_1\cdots\!\! \Int \!\!dy_n \Int \!\!dx\!\!\Int\!\!dv\>
e^{(xk + v\th)}\,\Wn \,\dx \dv\, . \eqno(2.9)
$$
Performing the integration over $x$ and $v$ and replacing $x_n$ and $v_n$
with the corresponding expressions from Eqs.~(2.5) which relate
them to the values $y_i$, we have:
$$
\eqalignno{
\fpn = \Int\!\! dy_1\cdots\!\! \Int \!\! dy_n\>& e^{(u_nk +
w_n\th)} \Wn\cr
&\times \exp\left\{i\sum_{j=1}^n [k g(n,i)+\th
h(n,i)]y_i\dt\right\}\,. &(2.10)\cr
}
$$
Defining $\f_i = [g(n,i)k + h(n,i)\th]\dt$, and comparing with
Eq.~(2.8b), we see that
$$
\eqalignno{
\fpn =& e^{i(u_nk + w_n\th)}\wn \, ,&(2.11a) \cr
\noalign{\hbox {where}}
\f_i =& [g(n,i)k + h(n,i)\th]\dt\, .&(2.11b)\cr
}
$$
Ultimately, however, we are interested in the FT of $p(x,v,t)$,
$$
\fpt = \Int \!\!dx \Int\!\!dv \> e^{i(xk
+v\th)}\,p(x,v,t)\,.\eqno(2.12)
$$
To obtain this, we take the limit $\li$ of $\fpn$, \ie,
$$
\eqalignno{
\fpt =& \lim_{\scriptstyle {{{\scriptstyle n\to \infty} \atop
\scriptstyle \dt \to 0} \atop \scriptstyle n\dt = t} }
e^{i(u_nk + w_n\th)}\wn  \, , &(2.13a) \cr
\noalign{\hbox {with}}
\f_i =& [g(n,i)k + h(n,i)\th]\dt\, .&(2.13b)\cr
}
$$

Eq.~(2.13) is the central result of this section and the basis
for the rest of this paper, which deals with specific applications.

Because of its importance, one should note that while reasonable from
the physical point of view, Eq.~(2.13) nonetheless raises some subtle
mathematical questions. We cannot prove that the limit $\li$ of $\fpn$
exists in general, nor that given its existence, it is indeed equal to
$\fpt$. The source of this difficulty is Eq.~(2.7), which is clearly a
discrete path integral, and therefore subject to various mathematical
reservations when the limit $n \to \infty$ is taken formally.

However, in this paper the usual formalism associated with path
integrals is bypassed in favor of the direct definition Eq.~(2.13). In
other words, we do not take the limit $\li$ formally and write the
resulting expression as a path integral. Rather, all calculations are
carried out for the finite $n$ case, and the limit then taken directly.
It is remarkable that in many cases this limit turns out to be
well defined mathematically as well as calculable, with the only
assumptions being the smoothness properties usually postulated in
physical problems. The next two sections are devoted to such
calculations and cover a wide range of cases.

\bigskip\centerline{\bf III. Monovariant Markov Noise Processes }\smallskip

As a first example where the limit in Eq.~(2.13) can be calculated
explicitly, consider a class of Markov noise processes defined below:

A Markov process is fully determined by the initial distribution $\po$,
and the transition probability, $T(y_{i+1},t_i + \dt \vert y_i,t_i)$,
from the value $y_i$ of the random function $R(t_i)$ at time $t_i$ to
the value $y_{i+1} = R(t_i + \dt)$ at a time $t_i+\dt$ [12]. It is
usually assumed that this function depends only on the time step $\dt$
and not on $t_i$. I shall assume this for simplicity, but the
calculations can be modified for more general cases.

I shall say a Markov process is {\it monovariant} if the transition
probability does not depend on $y_{i+1}$ and $y_i$ separately, but rather
on a single linear combination of these variables. Such a
combination can always be written as $\yy$, where $\k (\dt)$ is some
function of the time step $\dt$, and I also assume that $\k(0) \neq 0$.
Hence,
$$
T(y_{i+1},\dt \vert y_i) = T_{\dt}(\yy )\, .\eqno(3.1)
$$

For any Markov process, the functions $\Wn$ are given by [12]:
$$
\Wn = \po \,T_{\dt}(y_2 \vert y_1)\,T_{\dt}(y_3 \vert y_2) \cdots
T_{\dt}(y_n\vert y_{n-1})\, ,\eqno(3.2)
$$
where we assume all transitions take place during a time-step $\dt
=t/n$, with $n$ eventually going to infinity.

Many of the noises used in physical problems are monovariant. Some
examples are the Wiener noise (with $\k (\dt) = 1$), the Cauchy noise
(also with $\k (\dt) = 1$), and the Ornstein-Uhlenbeck (O-U) noise, for
which  $\k (\dt) = \exp (-\dt/ \tau)$, where $\tau$ is the correlation
time [12]. The O-U noise is widely used to describe forces with finite
correlation time (\eg, Ref.~[9]). The $\tau \to 0$ limit corresponds to
white noise.

Let us now calculate the pdf for a process driven by a general
monovariant Markov noise: The multi-variable FT of $\Wn$ is
$$
\eqalignno{
\wn = \Int\!\! dy_1 \cdots \!\!\Int\!\! dy_n\,& \exp\left[i(y_1\f_1
+\cdots +y_n\f_n)\right]\,\po \cr
&\times T_{\dt}(y_2 \vert y_1)\,T_{\dt}(y_3 \vert y_2) \cdots
T_{\dt}(y_n \vert y_{n-1})\, . &(3.3)\cr
}
$$
If the Markov noise is monovariant, so that $T_{\dt}(y'\vert y) =
T_{\dt}\left(y'-\k(\dt)y\right)$, we can perform the integration in the
following way: Define a single-variable Fourier transform $\widehat T$
such that
$$
\widehat T_{\dt}(\xi) = \Int \!\! e^{i\xi\eta} \,T_{\dt}(\eta)\,
d\eta\, ,\eqno(3.4)
$$
where $\eta = y'- \k(\dt)y$ is the single variable on
which $T_{\dt}(y'\vert y)$ depends, except for any direct
dependence on $\dt$. For conciseness, I assume such a dependence
implicitly and do not write the $\dt$ subscript anymore. We now
change the integration variables $\{y_i\}_{i=1}^n$ in Eq.~(3.3) to the new
set:
$$
\vbox{\eqalignno{
\eta _1 &= y_1 \, , &(3.5)\cr
\eta _2 &= y_2 -\k (\dt)y_1 \, , \cr
\eta _3 &= y_3 -\k (\dt)y_2\, , \cr
 {}&\vdots\cr
\eta _n &= y_n-\k (\dt)y_{n-1}  \, . \cr
}}
$$
The Jacobian of this transformation is clearly unity, and each
$T_{\dt}(y_i\vert y_{i-1})$ depends on $\eta_i$ only. To rewrite
$\sum_{i=1}^n y_i \f_i$ in terms of the new variables, we invert
Eq.~(3.5), which gives:
$$
\vbox{\eqalignno{
y_1 &= \eta_1 \, , &(3.6)\cr
y_2 &= \eta_2 +\k (\dt)\eta_1\, , \cr
y_3 &= \eta_3 +\k (\dt)\eta_2 +\left[\k (\dt)\right]^2\eta_1\, , \cr
{}&\vdots\cr
y_i &= \sum_{k=1}^i \left[\k (\dt)\right]^{k-1}\eta_{i-k+1} \, .\cr
}}
$$
Hence,
$$
\sum_{i=1}^n y_i\f_i= \sum_{i=1}^n \eta_i \left\{ \sum_{k=i}^n \f_k
\left[\k(\dt)\right]^{k-i}\right\} \, .\eqno(3.7)
$$
Substituting Eq.~(3.7) into the expression for $\widehat W$, Eq.~(3.30),
and using the definition of $\widehat T$, Eq.~(3.4), we finally obtain
$$
\eqalignno{
\wn &= \widehat P_0 \left[\zeta\right] \prod_{i=2}^n \widehat
T\left( \xi_i\right)  \, , &(3.8a)\cr
\noalign{\hbox{with}}
\zeta & \equiv \sum_{k=1}^n \f_k \left[\k(\dt)\right]^{k-1} \, ,
&(3.8b)\cr
\xi_i &\equiv \sum_{k=i}^n \f_k \left[\k(\dt)\right]^{k-i} \, .
&(3.8c)\cr
}
$$
In Eq.~(3.8a),
$$
\widehat P_0[\zeta] = \Int \!\!e^{i\zeta y} P_0(y)\,dy\, . \eqno(3.9)
$$

In accordance with the general formula (2.11), $\fpn$, is obtained by
replacing $\f_i$ in Eqs.~(3.11) with $[g(n,i)k + h(n,i)\th]\dt$. This
yields:
$$
\eqalignno{
&\fpn = e^{i(u_nk + w_n\th)}\widehat P_0
\left[\zeta\right]\prod_{i=2}^n \widehat T\left( \xi_i\right)\, ,
&(3.10a)\cr
\noalign{\hbox{where}}
&\zeta  \equiv k\left(\sum_{k=1}^n
g(n,k)\left[\k(\dt)\right]^{k-1}\dt\right) + \th\left(\sum_{k=1}^n
h(n,k) \left[\k(\dt)\right]^{k-1}\dt\right) \, , &(3.10b)\cr
&\xi_i \equiv k\left(\sum_{k=i}^n g(n,k)
\left[\k(\dt)\right]^{k-i}\dt\right)+\th\left(\sum_{k=i}^n h(n,k)
\left[\k(\dt)\right]^{k-i}\dt\right)\, . &(3.10c) \cr
}
$$

Last, we must take the limit $\li$. This is a somewhat tedious step, and
its details can be found in Appendix A. The final result is the
following very general formula:

$$
\eqalignno{
\fpt &= e^{i\left[ k u(t) + \th w(t) \right]}\widehat P_0 \left[
\zeta (t) \right] \exp \left\{ \int_0^t dt' B\left[ \xi (t,t') \right]
\right\} \, ,&(3.11a)\cr
\noalign{\hbox{where}}
B(\xi) &= \td {} \dt \left[ \log \widehat T_{\dt} (\xi)
\right]_{\dt=0}\, ,&(3.11b)\cr
\zeta (t) &\equiv k \int_0^t g(t,s)\Psi(s)\,ds + \th \int_0^t h(t,s)
\Psi(s)\, ds\, ,&(3.11c)\cr
\xi (t,t') &\equiv k \int_{t'}^t g(t,s)\Psi(s-t')\, ds\, + \,\th
\int_{t'}^t h(t,s) \Psi(s-t') \,ds\, ,&(3.11d)\cr
\noalign{\hbox{with}}
\Psi(z) &= \kappa (0)\exp[az]\, ,&(3.11e)\cr
a &\equiv \td {[\log \k(t)]} t \bigg\vert_{t=0} = {1 \over \k(t)}\td
{\k(t)} t \bigg\vert_{t=0} \, .&(3.11f)\cr
}
$$

Eqs.~(3.11) are general and can be applied for a great variety of noise
processes and of dynamics (\ie, forms of the Langevin equation). The
price of this flexibility is the complex form of the equations. Their
meaning and use can be clarified, however, by using as an example a very
simple equation which was investigated recently [8,9],
$$
\ddot x = \dot v = R(t)\, ,\eqno(3.12)
$$

The solution is $v=v_0 + \int_0^t R(t')dt'$, so that we have (compare
Eq.~(2.5)):
$$
\eqalignno{
w(t)&=v_0 \, , &(3.13a)\cr
\htt& = 1\, . &(3.13b)\cr}
$$
Now $x(t) = x_0 + \int_0^t dt''\, v(t'')= x_0 +v_0t + \int_0^t dt'
\int_0^{t'} dt''\, R(t'') $. According to a well-known theorem [13],
$$
\int_0^t dt'\int_0^{t'} \!\! dt''\>R(t'')=\int_0^t dt'\>(t-t')
R(t')\, ,\eqno(3.14)
$$
so that
$$
\eqalignno{
u(t)&=x_0+v_0 t \, , &(3.15)\cr
\gtt &= t-t'   \qquad\quad  (t'<t)\, . \cr}
$$
Let us now take $R(t)$ to be an O-U noise, defined by [12, 9]:
$$
\eqalignno{
\po&=\sqrt{\tau \over \pi f_0^2}\exp\left(-{\tau \over f_0^2} y^2\right)
\, .&(3.16a)\cr
T_{\dt}(y_{i+1}\vert y_i) &= \sqrt{\tau \over \pi f_0^2
\left[1-\k^2(\dt)\right]}\exp\left\{- {\tau\left[ (\yy)\right]^2 \over
f_0^2\left[1-\k^2 (\dt)\right]} \right\} \, .&(3.16b)\cr
\k(\dt)&=\exp\left(-{\dt \over \tau}\right)\, .&(3.16c)\cr
}
$$
where $f_0$ is a parameter describing the intensity of the random force.

The Fourier transforms $\widehat P_0$ and $\widehat T$, are (see
Eqs.~(3.4) and (3.9))
$$
\eqalignno{
\widehat P_0 [\zeta]& = \exp\left(-{f_0^2 \over 4\tau} \zeta^2\right)
\, .&(3.17a)\cr
\widehat T_{\dt}(\xi) &= \exp\left[{(1-\k^2)f_0^2 \over
4\tau }\xi^2\right]\, .&(3.17b)\cr
}
$$
{}From Eq.~(3.11f), we find
$$
a = - {1 \over \tau}\, ,\eqno(3.18)
$$
so that $\Psi(z)$ is now (from Eq.~(3.11e))
$$
\Psi(z) = \exp\left(-{z \over \tau}\right)\, .\eqno(3.19)
$$
{}From the definition of $B(\xi)$, Eq.~(3.11b), we have:
$$
B(\xi) = -{f_0^2 \over 2\tau} \xi^2\, .\eqno(3.20)
$$
The details of the remaining algebra are in Appendix A. The final result
is
$$
\eqalignno{
\hat p(k ,\th ,t)& = \exp\left\{i\left[k(x_0+v_0t) + \th
v_0\right]-\h\left[\alpha(t)k^2 +2\beta(t)k\th + \gamma(t)\th^2
\right]\right\}\,, &(3.21a)\cr
\noalign{\hbox{where}}
\alpha(t)&=f_0^2\left[{t^3 \over 3} -\h t^2\tau
+\tau^3\left(1-e^{-t/\tau}\right) -\tau^2te^{-t/\tau}\right]\, .
&(3.21b)\cr
\beta(t)&=\h f_0^2\left[t^2-\tau
t\left(1-e^{-t/\tau}\right)\right]\, .&(3.21c)\cr
\gamma(t)&=f_0^2\left[t-\tau\left(1-e^{-t/\tau}\right)\right]\,
.&(3.21d)\cr
}
$$

Since this is a Gaussian Fourier transform, it can be inverted easily to
yield the real space joint pdf:
$$
\vbox{\eqalignno{
p(x,v,t)&= {1 \over 2\pi \sqrt{(\alpha \gamma - \beta^2)}}\cr
\times &\exp\left\{-{1\over 2(\alpha \gamma - \beta^2)}\left[\gamma
(x-x_0-v_0t)^2 -2 \beta (x-x_0-v_0t)(v-v_0) +
\alpha(v-v_0)^2\right]\right\}\, ,\cr
&{} &(3.22)\cr
}}
$$
with $\alpha, \beta$ and $\gamma$ as defined in Eq.~(3.21).

The marginal distributions for the position, $p(x,t)$, and for the
velocity, $p(v,t)$, are most easily obtained from Eq.~(3.21) by setting
$k=0$ and $\th=0$ respectively, then Fourier inverting. The results
are:
$$
\eqalignno{
p(x,t)&= {1 \over \sqrt{2\pi \alpha (t)}}\exp\left[-{(x-x_0-v_0t)^2 \over
2\alpha (t)}\right] \, .&(3.22a)\cr
p(v,t)&= {1 \over \sqrt{2\pi \gamma (t)}}\exp\left[-{(v-v_0)^2 \over
2\gamma (t)}\right]\, .&(3.22b)\cr
}
$$

Eq.~(3.11) with an O-U noise was investigated by Heinrichs [9], who
derived Fokker-Planck (FP) equations for the joint pdf as well as for
the two marginal pdf 's $p(x,t)$ and $p(v,t)$. He solved these
equations approximately only in two limiting cases: For very small
$\tau$ (near $\tau=0$, which corresponds to white noise) and for very
long $\tau$ (near $1/{\tau}=0$, which correspond to a constant force).

Here, however, we have obtained exact results in Eqs.~(3.22) and (3.23),
quite easily and without solving any differential equation. These
expressions reduce to Heinrichs's results in the appropriate
limits. Also, a simple substitution shows that Eqs.~(3.22)-(3.23) are
indeed solutions of Heinrichs's FP equations [9]. This further confirms
the effectiveness of the present approach.

\bigskip\centerline{\bf IV. Gaussian Processes}\smallskip

Another class of random noises that can be solved completely is the
group of Gaussian noises. Except for the O-U noise,
all these processes are non-Markovian, and they are defined by the
requirement that all the distribution functions $\Wn$ be of Gaussian
form. For zero-mean noises, this is equivalent (\eg, Ref.~[7]) to the
requirement that their Fourier transform be given by:
$$
\wn=\exp\left[-\h \sum^n_{i,j = 1} \f_i\f_j\p(i,j)\right]\,
,\eqno(4.1)
$$
where $\p(i,j)$ is the correlation function of the noise at time steps
$i$ and $j$, \ie, $\av {y_iy_j}$. Note that formula (4.1) assumes that
$\p(i,j)=\p(j,i)$.

Substituting (4.1) into (2.11), we have:
$$
\eqalignno{
\fpn = \exp\Biggl\{i(k u_n +\th w_n) - \h \Biggl[&k^2 \sum_{i,j=0}^n
g(n,i)g(n,j)\p(i,j)\,\,\dt^2\cr
+ &k\th \sum_{i,j=0}^n \left[g(n,i)h(n,j)+
h(n,i)g(n,j)\right]\p(i,j)\,\,\dt^2\cr
+& \th^2 \sum_{i,j=0}^n h(n,i)h(n,j)\p(i,j)\,\,\dt^2
\biggr]\Biggr\}\, .&(4.2)\cr
}
$$
Upon taking the limit $\li$, all the sums
become integrals. Hence, we have from Eq.~(2.13):
$$
\eqalignno{
\fpt& =  \exp\left\{i(k u(t) +\th w(t)) - \h \left[\alpha (t)k^2
+2\beta (t) k\th + \gamma (t) \th^2\right]\right\} \,,
&(4.3a)\cr
\noalign{\hbox{where}}
\alpha (t)&= \int_0^t \!\!dt'\!\! \int_0^t\!\! dt''\, g(t,t')g(t,t'')
\,\p(t',t'')\, ,&(4.3b)\cr
\beta (t)&= \h \int_0^t\!\! dt'\!\! \int_0^t\!\! dt''\,
\left[g(t,t')h(t,t'')+ h(t,t')g(t,t'')\right]\, \p (t',t'') \, ,
&(4.3c)\cr
\gamma (t)&= \int_0^t\!\! dt' \!\!\int_0^t \!\!dt''\, h(t,t')h(t,t'')
\,\p (t',t'')\, , &(4.3d)\cr
}
$$
and $\p(t',t'') =\av {R(t')R(t'')}$ is the continuous time correlation
function of the noise $R(t)$.

As an application of Eq.~(4.3) and as a consistency check, we can use
Eq.~(3.12) again, $\ddot x = R(t)$, with $R(t)$ as O-U noise (since this
noise is both Markovian and Gaussian). For this case
$$
\eqalignno{
\p(t',t'') &= {f_0^2 \over {2\tau}}\exp\left[- {\left\vert {t'-t''}
\right\vert \over \tau}\right] \, ,&(4.4a)\cr
g(t,t') &= t-t'    \qquad\qquad\qquad\,  (t>t')\, ,&(4.4b)\cr
h(t,t') &=1\, .&(4.4c)\cr
}
$$

The functions $\alpha, \beta$ and
$\gamma$ calculated through Eq.~(4.3) are identical (as expected) with
those obtained for the same noise (O-U) in Eq.~(3.21).

\bigskip\centerline{\bf V. Differential Equations for PDF's in the
Gaussian Case}\smallskip

The preceding sections presented cases where direct calculation of the
pdf could be performed, \ie, when the limit in Eq.~(2.13) could be
calculated explicitly. In some cases, however, the limit proves too
complicated for explicit calculation. Eq.~(2.11) then offers an
alternative point of view by providing a general approach for deriving
FP-like differential equations for the pdf's.

To exemplify this approach through a simple example, consider the
Gaussian random noises described in section IV. Having arrived at
Eq.~(4.2), we now choose not to take the limit $\li$ which would yield
Eqs.~(4.3). Instead, let us calculate the time derivative of $\fpt$ from
its discrete form $\fpn$. This is just
$$
\pd {\fpt} t = \lim_{\dt \to 0} { {\hat p(k,\th,n+1) -\fpn} \over
\dt}\equiv \lim_{\dt \to 0} {\Delta \hat p \over \dt} \, .\eqno(5.1)
$$

With the help of Eq.~(4.2) for $\fpn$, this calculation is
straightforward but tedious, and the details can be found in Appendix
B. The final result is:
$$
\pd {\fpt} t =  \left\{i \td u t k +i\td w t \th -\h \left[
\td \alpha t k^2 +  2 \td \beta t k\th + \td \gamma t \th^2
\right]\right\}\fpt\, ,\eqno(5.2)
$$
where $\alpha, \beta$ and $\gamma$ are the functions defined in
Eqs.~(4.3).

Eq.~(5.2) can be Fourier inverted to yield:
$$
\pd {p(x,v,t)} t = -\td u t \pd p x -\td w t \pd p v +\h \left[ \td
\alpha t {\pdd  p x} +  2\td \beta t {\partial^2 p
\over {\partial x \partial v}} + \td \gamma t {\pdd p v}\right]\,
.\eqno(5.3)
$$
Here again, one should note the high generality of these equations. The
result obtained here is valid for a wide range of Langevin equations as
well as for a wide range of random noises.

We can now apply this formalism to the simple equation $\ddot x = R(t)$
with the random force being the O-U noise already considered in section
III  for which we have found the pdf [Eq.~(3.22)]. Then from (4.3)
and (4.4), we have
$$
\eqalignno{
\td \gamma t& = f_0^2\left[ 1- e^{-t/\tau}\right] \, .&(5.4a)\cr
\td \beta t &= \h f_0^2\left[ t(2 - e^{-t/\tau})-\tau(1 -
e^{-t/\tau})\right]\, .&(5.4b)\cr
\td \alpha t &= f_0^2\left[ t^2- t\tau(1 - e^{-t/\tau})\right]\,
.&(5.4c)\cr
}
$$

The remarkable thing about Eqs.~(5.3)-(5.4) is that they {\it differ}
from the FP equations derived by Heinrichs for the {\it same} stochastic
process [9]. Instead of Eqs.~(5.3)-(5.4), Heinrichs obtains (for the
case he investigated, \ie, $x_0=v_0=0$, hence $u=w=0$) :
$$
\eqalignno{
\pd {p(x,v,t)} t &= \left[-v \pd {} x -b(t) {\partial^2 {}\over {\partial
x \partial v}} + a(t) {\pdd  {} v} \right]
p(x,v,t) \, ,&(5.5a)\cr
\noalign{\hbox{where}}
a(t)&= {f_0^2\over 2}\left[1- e^{-t/\tau}\right]\, ,&(5.5b)\cr
b(t)&={f_0^2\over 2}\left[\left(t+\tau\right)\,e^{-t/\tau} -
\tau\right]\, .&(5.5c)\cr
}
$$
This equation is not wrong. Indeed, as mentioned before, the function
$p(x,v,t)$ calculated in Eq.~(3.29) {\it is} the solution of Heinrichs's
equation (5.5), {\it as well} as the solution of Eq.~(5.3) (both with
the appropriate initial conditions). This means that the FP equation
corresponding to a given Langevin equation is {\it not unique} (and the
existence of two different equations means trivially that there is an
infinite number of them). Rather, the form of the equation depends on
the specific method used to derive it. All the various FP equations,
however, are equivalent to each other through the specific mathematical
form of the solution.

It may seem that this last observation, as well as having the explicit
solution, renders all discussions about the equations unimportant. This
may be true for the case of the unconstrained particle, which is discussed
here, but the discussion is very relevant to cases of constrained
particles.

A particle can be ``constrained'', \eg, by the addition of
boundaries to the system (for example, absorbing boundaries are often
added for calculating first passage times). In the theory of the
classical FP equation [14], the presence of such boundaries doesn't
change the equation. Rather, it implies new mathematical boundary
conditions on $p(x,v,t)$ (these usually make the task of solving the
equations much more difficult).

It is tempting to extend such reasoning to Fokker-Planck-{\it like}
equations, such as Eqs.~(5.3) or (5.5). Indeed, Heinrichs used
precisely this approach to tackle the case of a particle driven by O-U
noise when two absorbing boundaries are present [15]. However, the
non-uniqueness of the FP-equation casts serious doubt on the correctness
of such an approach. It is highly unlikely, for example, that Eq.~(5.3)
and (5.5) will still be equivalent to each other when new boundary
conditions are added to them. Furthermore, it is quite possible that
neither of them is correct in the presence of boundaries. Indeed,
Eq.~(5.3) follows from the general Eq.~(2.11), which assumes implicitly
the absence of constraints in so far as it allows summation over {\it
all} paths that lead to the adequate final position and velocity. On the
other hand, Heinrichs derived Eq.~(5.5) by applying some averaging
identities to the stochastic differential equation (3.12) [9]. Under
constraints, such averaging procedures may change and one cannot assume
\apr \ that the FP equation remains the same (or indeed exists at all).

In other words, the non-uniqueness of the FP-equations implies that
whenever we add boundaries to the problem, we must provide independent
mathematical justification for any FP equation we choose to use. In the
absence of such justification, the results obtained by adding boundary
conditions to any FP equations cannot be trusted.

\bigskip\centerline{\bf VI. Differential Equations for PDF's in the
Case of Markovian Dichotomous Noise}\smallskip

For Gaussian noises, we have both an explicit expression for the pdf
and a FP-like equation Eq.~(5.3). However, and even more importantly,
the method described in the last section can be used in cases where the
limit $\li$ proves too complex to perform explicitly. As an example of
the usefulness of this approach, I discuss now a specific type of random
driving force, the Markovian dichotomous noise, which is used, for
example, in the theory of second order Butterworth filters [16].

Masoliver [10,11] has treated two stochastic differential equations
with this particular noise, \ie,  $\ddot x = R(t)$ and $\ddot x +
\gamma\dot x = R(t)$, and has obtained FP-like equations for $p(x,v,t)$,
$p(x,t)$ and $p(v,t)$ in both cases. Deriving the equations requires
much work in each case, yet the general formula (2.13) provides a
unifying point of view from which all these equations follow
naturally. In this section, I derive a generic differential equation for
all the various pdf's, when the driving force is the Markovian
dichotomous noise, but the form of the equation is still
general. Masoliver's equations follow immediately as several particular
cases of this general equation. Thus, we shall obtain a powerful
generalization as well as a compact form for all the cases studied.

The Markovian dichotomous noise is defined by:
$$
\eqalignno{
\po &= \h \left[\delta(y_1-a) + \delta (y_1+a)\right] \, ,&(6.1a)\cr
T_{\dt}(y \vert y') &= \h \left[f(\dt)\delta (y-y') +g(\dt)\delta
(y+y')\right]\, ,&(6.1b)\cr
\noalign{\hbox{where}}
f(\dt) &= 1+ \exp\left(-2\lambda\dt\right)\, ,&(6.1c)\cr
g(\dt) &= 1- \exp\left(-2\lambda\dt\right)\, .&(6.1d)\cr
}
$$
Thus the force alternates between two values only, $a$ and $-a$, and the
probability that a switch from one value to the other occurs in the
interval $(t,t+dt)$ is $\lambda e^{-\lambda t}\, dt$ [10]. The parameter
$\lambda$ is the average time between switches.

The calculation of $\fpn$ follows the lines described in sections
II-IV. The details are in appendix C. The final result is
$$
\vbox{\eqalignno{
\fpn = \exp\left[i(k u_n +\th w_n)\right]
\sum_{k=1}^n &\left(\h\right)^n f^{n-k-1}g^{k-1}\cr
&\times  \left[F^n_k(\f_1,
\ldots, \f_n) + F^n_k(-\f_1, \ldots, -\f_n)\right] \, , &(6.2a)\cr
\noalign{\hbox{where}}
\f_i=\left[k g(n,i)+\th h(n,i)\right]\dt& \, .&(6.2b)\cr
} }
$$
The coefficients $F^n_k(\f_1, \ldots, \f_n)$ are determined recursively
by
$$
\eqalignno{
F^n_k(\f_1,\ldots, \f_n)& = e^{ia\f_1}\left[F^{n-1}_k(\f_2,\ldots,\f_n)
+F^{n-1}_{k-1}(-\f_2,\ldots,-\f_n)\right] \, ,&(6.3a)\cr
\noalign{\hbox{with the conditions}}
F^1_1(\f_1)&=e^{ia\f_1}\, .&(6.3b)\cr
F^m_0(\f_1,\ldots, \f_n)&=0\, .&(6.3c)\cr
F^m_{m+1}(\f_1,\ldots, \f_n)&=0 \, .&(6.3d)\cr
}
$$
In principle, Eq.~(6.3) is the full solution of the problem. However,
the limit $\li$ cannot be taken explicitly in all
generality. Nonetheless, Eq.~(6.3) provides an approximation of $\fpt$,
which can be calculated by selecting a sufficiently high value of $n$.

In the following, however, I use Eq.~(6.3) for deriving the
differential equation obeyed by $\fpt$ along the lines of section V. In
the present case, however, this procedure yields a  second order
equation in the time derivatives, rather than a first order one as in
the case of Gaussian noise.

For the purpose of this derivation, I shall assume that $g(n,i)$ and
$h(n,i)$ depend on the difference $n-i$ only (or, in the continuum
version $\gtt=g(t-t')$). This is usually true. Also, I assume that
$u(t)=w(t)=0$ (the particle starts at rest from the origin). This,
however, is no loss of generality, because if we are interested in other
cases, all we need to do is multiply the function $\fpt$ obtained for
the case $u(t)=w(t)=0$ by $\exp[i(k u(t) + \th w(t)]$.

The details of the calculations can be found in appendix D. The main
results are:

There is no first order equation in the time derivatives. This is
because, to fist order in $\dt$, we find that
$$
\Delta \fpn = ia\f_1 \sum_{k=1}^n\left(\h\right)^n\!\!
f^{n-k-1}g^{k-1} \left[F^n_k(\f_2, \ldots, \f_{n+1}) - F^n_k(-\f_2,
\ldots, -\f_{n+1})\right] \, ,\eqno(6.4)
$$
with $\f_i$ having the same meaning as in Eq.~(6.2), \ie,
$\f_i=\left[k g(n,i)+\th h(n,i)\right]\dt$. Comparing this expression
with Eq.~(6.2) reveals that $\Delta \hat p$ cannot be expressed in terms
of $\fpn$ only, hence we cannot write down a first order equation.

We therefore look at the second derivative, for which we
need to calculate $\Delta^2 \hat p \equiv \hat p(k,\th,n+2) + \fpn
-2\hat p(k,\th,n+1)$. This expression is then expanded to second
order in powers of $\dt$. As shown in appendix D, we can then express
$\Delta^2 \hat p / \dt^2$ in terms of $\Delta \hat p / \dt$ and $\hat
p$, so that in the limit $\li$ we obtain a second order differential
equation for $\fpn$. Quoting from appendix D, we have finally that
$$
\eqalignno{
&\pdd {\fpt} t +\left(2\lambda - {1\over \f} \td \f t\right)\pd {\fpt} t
+ a^2\f^2\fpt = 0  \, ,&(6.5a)\cr
\noalign{\hbox{where}}
&\f \equiv k g(t) + \th h(t)\, . &(6.5b)\cr
}
$$

The power of Eq.~(6.5) lies in the generality of the functions $g(t)$
and $h(t)$ that appear in it. This is because Eq.~(6.5) covers all
linear stochastic processes driven by a Markovian dichotomous noise.
Thus, although the noise is fully determined, the dynamics remains
arbitrary [within the confines set by Eq.~(2.2)], and therefore,
Eq.~(6.5) covers at once a wide range of stochastic processes.

Consider, for example, the simple equation $\ddot x = R(t)$, for which
$g(t)=t, h(t)=1$. Then , from Eq.~(6.5) we have immediately for $\fpt$:
$$
\pdd {\fpt} t +\left(2\lambda - {k \over k t + \th}\right)\pd
{\fpt} t + a^2\left(k t + \th\right)^2\fpt = 0\, .\eqno(6.6)
$$
The equations for the FT of the marginal distribution for the position,
\ie, $\hat p(k,t)$ and for the velocity, \ie, $\hat p(\th,t)$ follow at
once by setting $\th=0$ and $k=0$ respectively:
$$
\eqalignno{
\pdd {\hat p(k,t)} t +\left(2\lambda - {1 \over t}\right)\pd {\hat
p(k,t)} t + a^2k ^2 t^2 \,\hat p(k,t) &= 0 \, . &(6.7a)\cr
\pdd {\hat p(\th,t)} t + 2\lambda \pd {\hat p(\th,t)} t +
a^2\th ^2 \,\hat p(\th,t) &= 0 \, . &(6.7b)\cr
}
$$
Eq.~(6.7b) is the FT of the telegraph equation for $p(v,t)$. The two
Eqs.~(6.7) were derived separately by Masoliver [10], along with a
somewhat more complicated version of Eq.~(6.7) (the complication being
due mainly to notations). As Masoliver pointed out, $k$ and $\th$ do
not appear independently in the equation for $\fpt$, but do so only
through the combination $k t + \th$. This fact was not immediately
apparent in his version of the equation for $\fpt$, but it was used
later to derive Eq.~(6.6).

There is no obvious reason in Masoliver's work for the appearance of
such a combination. Here, on the other hand, it appears as a natural
consequence of the general formula, Eq.~(2.13), in which $k$ and $\th$
are always bound in a well defined combination (except where they relate
to $u(t)$ and $w(t)$). This also makes very clear the meaning of the
particular coefficients in this combination ($t$ and 1), as they are
merely the Green's functions of the original stochastic equation.

Masoliver later tackled the more general equation  $\ddot x+\gamma\dot x
= R(t)$ [11]. Much work is needed to derive the FP equations for $\fpt$,
$\hat p(k,t)$ and $\hat p(\th,t)$. Moreover, in Masoliver's approach,
the equations for the marginal distributions do not follow immediately
from the equation for $\fpt$, but instead require still more work. In the
present approach, on the other hand, the derivations are almost trivial.
For the equation  $\ddot x+\gamma\dot x = R(t)$, we have
$$
\eqalignno{
g(t-t')  &={1 \over \gamma}\left\{1 - \exp\left[-\gamma
(t-t')\right]\right\} \, .&(6.8a)\cr
h(t-t') &= \exp[-\gamma (t-t')]\, .&(6.8b)\cr
}
$$
(easily derived, \eg, from Laplace transforming the equation). Thus, we
have at once:
$$
\eqalignno{
\pdd {\fpt} t +\left[2\lambda - {\gamma(k- \gamma \th)e^{-\gamma t}
\over k -(k- \gamma \th)e^{-\gamma t}}\right]\pd {\fpt}
t\qquad\qquad\cr
 +{a^2\over \gamma^2}\left[k - (k- \gamma \th)e^{-\gamma
t}\right]^2 \fpt& = 0 \, .&(6.9a)\cr
\noalign{\hbox{}}
\pdd {\hat p(k,t)} t +\left[2\lambda - {\gamma e^{-\gamma t} \over 1 -
e^{-\gamma t}}\right]\pd {\hat p(k,t)} t + {a^2k^2\over
\gamma^2}\left(1 - e^{-\gamma t}\right)^2 \hat p(k,t)& = 0 \,
.&(6.9b)\cr
\noalign{\hbox{}}
\pdd {\hat p(\th,t)} t +(2\lambda + \gamma)\pd {\hat p(\th,t)} t +
a^2\th^2 e^{-2\gamma t} \hat p(\th,t)& = 0 \, .&(6.9c)\cr
}
$$
These agree with Masoliver's equations, but are derived much more
easily. Again, the fact that $k$ and $\th$ appear only through the
combination $k \left[1 - \exp(-\gamma t)\right] + \th
\gamma\exp(-\gamma t)$ is not evident in Masoliver's approach, while it
finds a very natural explanation in the present framework.

\bigskip\centerline{\bf VII. Summary}\smallskip

The main thrust of this work is methodological and conceptual. Rather
than concentrating on a specific process, I have presented
a general approach to a large class of stochastic differential equations.

The basis of the method is the discretization of the process described
by the equation into $n$ time-steps. Each realization of the process
consists then of $n$ steps, and its probability can be calculated from
the elementary properties of the random noise. In this discrete
approximation, the probability of finding the
particle in a given state at time $t$ is the sum of the probabilities of
the realizations which lead to this final state. Taking $n \to \infty$
then yields the required probability for the continuous process.

This method is successful on at least two fronts. First, if the limit $n
\to \infty$ can be calculated explicitly, we obtain a closed expression
for the required probability distribution function. This turns out to be
the case for all Gaussian processes and monovariant Markov
processes. There is no reason to believe these exhaust the possibilities
of this method. Quite likely, other processes can be completely
solved in this way.

Second, the expressions obtained for the discrete approximation are
useful for deriving Fokker-Planck equations for the probability
distribution functions. For the case of the Markovian dichotomous noise,
I have shown that there is a generic second order FP equation, which
covers all the possible stochastic differential equations to which
the general method is applicable. Thus, we obtain a unified
point of view on the various FP equations this particular noise can
generate. In particular, for a joint probability distribution involving
more than one stochastic variable (\eg, position and velocity), the
Fourier frequencies corresponding to the stochastic  variables appear
only in a well specified linear combination. The coefficients in this
combination are the Green's function attached to each stochastic
variable. Thus, an important mathematical property of the FP equation is
made explicit and its origin is clarified. This was not obvious
in other approaches which have been applied to this problem.

Another general point underscored by the present approach is that there
may be several FP equations corresponding to a single stochastic
differential equation. This non-uniqueness may have important
implications when we consider particles  constrained, \eg, by various
boundaries. The standard recipe in Fokker-Planck theory, \ie, adding some
boundary conditions, is incomplete, since the various FP equations
may yield different results when supplemented with the {\it same}
boundary conditions. Thus, we must be careful when applying FP equations
derived for free particles to constrained cases. Justification is always
needed when taking such a step.

Finally, the discrete approximation should be of interest for numerical
estimates of the pdf. I have not tried to address this issue here, and
concentrated on analytical applications instead, but this should be
investigated by interested parties.

In the last few years, there seem to be somewhat more interest in
stochastic equations with complex environments and simple noises than
 in equations with simple environments and complex noises. I feel at
least part of the reason for this is that the latter cases have proved
quite difficult mathematically. Certainly, many physical systems do
exhibit complex random noises. The method presented here may simplify
the mathematical difficulties to a considerable extent. I hope this will
motivate interested researchers to use it and develop it further.

\bigskip
\centerline{\bf Acknowledgments}\smallskip

Many thanks to S. Redner, P. Krapivsky and E. Ben Naim for very
instructive discussions and criticisms, and to C. Doering for his
suggestions and the references he contributed.

\vfill\eject
\centerline {\bf Appendix A}\smallskip

We wish to take the limit $\li$ of Eq.~(3.10), \ie,
$$
\eqalignno{
&\fpn = e^{i(u_nk + w_n\th)}\widehat P_0
\left[\zeta\right]\prod_{i=2}^n \widehat T\left( \xi_i\right)\, ,
&(A.1a)\cr
\noalign{\hbox{where}}
&\zeta  \equiv k\left(\sum_{k=1}^n
g(n,k)\left[\k(\dt)\right]^{k-1}\dt\right) + \th\left(\sum_{k=1}^n
h(n,k) \left[\k(\dt)\right]^{k-1}\dt\right) \, , &(A.1b)\cr
&\xi_i \equiv k\left(\sum_{k=i}^n g(n,k)
\left[\k(\dt)\right]^{k-i}\dt\right)+\th\left(\sum_{k=i}^n h(n,k)
\left[\k(\dt)\right]^{k-i}\dt\right)\, . &(A.1c) \cr
}
$$

To make this step clearer, I'll perform it in several stages.

First, the expression $\left[\k(\dt)\right]^{k-i}$ seems to become
ill-defined in this limit. Let us therefore introduce the following
quantity:
$$
a \equiv \td {[\log \k(t)]} t \bigg\vert_{t=0} = {1 \over \k(t)}\td
{\k(t)} t \bigg\vert_{t=0} \, .\eqno(A.2)
$$
Hence, to first order in $\dt$,
$$
\k(\dt) \approx \k(0)\exp(a\dt)\, ,\eqno(A.3)
$$
so that
$$
\left[ \k(\dt) \right] ^{k-i} \approx \k(0)\exp \left[ a(k-i)\dt \right]
\equiv \Psi (k-i)\, .\eqno(A.4)
$$
This expression has a well defined value in the limit $\li$, when the
function $\Psi$ becomes
$$
\Psi (z) = \k(0) \exp(az)\, . \eqno(A.5)
$$

Next, note that all sums of the form $\sum_{k=i}^n \cdots \dt$ will
become integrals in this limit. Introducing an integration variable $s
\equiv k\dt $, an initial time $t' \equiv i\dt$ and the final time $t
\equiv n\dt$, and using the definition of $\Psi$, Eq.~(A.5), we have:
$$
\sum_{k=i}^n g(n,k) \left[ \k(\dt) \right] ^{k-i}\dt \longrightarrow
\int_{t'}^t g(t,s)\, \Psi (s-t')\, ds\, .\eqno(A.6)
$$
Similarly,
$$
\eqalignno{
\sum_{k=i}^n h(n,k) \left[ \k(\dt) \right]^{k-i}\dt &\longrightarrow
\int_{t'}^t  h(t,s)\, \Psi (s-t') \, ds \, , &(A.7a)\cr
\sum_{k=1}^n g(n,k) \left[ \k(\dt) \right]^{k-1}\dt &\longrightarrow
\int_{0}^t g(t,s) \,\Psi (s)\, ds\, , &(A.7b)\cr
\sum_{k=1}^n h(n,k) \left[ \k(\dt) \right]^{k-1}\dt &\longrightarrow
\int_{0}^t h(t,s) \,\Psi (s)\, ds \, , &(A.7c)\cr
}
$$
where in the last two lines, the difference between $(k-1)\dt$ and
$k\dt$ is neglected, since it yields a vanishing correction of order
$\dt$ to the expressions on the r.h.s.

Finally, consider the product over all the $T$-functions in Eq.~(A.1a).
Following the idea used in Eqs.~(A.2)-(A.5), we define
$$
B(\xi) = \td {} \dt \left[ \log \widehat T_{\dt} (\xi)
\right]_{\dt=0}\, .\eqno(A.8)
$$
{}From the definition of $\widehat T_{\dt}(\xi)$, we have $\widehat
T_{\dt}(0) = 1$. Hence, in the limit $\li$,
$$
\widehat T_{\dt}(\xi) \approx \exp\left[ B(\xi)\dt\right]\,
.\eqno(A.9)
$$

Substituting Eq.~(A.9) into the expression $\prod_{i=2}^n \widehat
T\left( \xi_i\right)$ which appears in Eq.~(A.1a) yields
$$
\prod_{i=2}^n \widehat T\left( \xi_i\right) \approx
\exp\left[\sum_{i=2}^n B(\xi_i)\dt \right]\,.\eqno(A.10)
$$
In the limit $\li$, we have, from Eq.~(A.1c), that
$$
\xi_i \longrightarrow \xi (t,t') \equiv k
\int_{t'}^t g(t,s)\Psi(s-t')\, ds\, + \,\th \int_{t'}^t h(t,s)
\Psi(s-t') \,ds\, , \eqno(A.11)
$$
and therefore
$$
\sum_{i=2}^n B(\xi_i)\dt  \longrightarrow \int_0^t dt' B\left[ \xi
(t,t') \right] \, .\eqno(A.12)
$$

Putting together Eqs.~(A.1), (A.6), (A.7) and (A.12), we finally have
that
$$
\eqalignno{
\fpt &= e^{i\left[ k u(t) + \th w(t) \right]}\widehat P_0 \left[
\zeta (t) \right] \exp \left\{ \int_0^t dt' B\left[ \xi (t,t') \right]
\right\} \, ,&(A.13a)\cr
\noalign{\hbox{where}}
\zeta (t) &\equiv k \int_0^t g(t,s)\Psi(s)\,ds + \th \int_0^t h(t,s)
\Psi(s)\, ds\, ,&(A.13b)\cr
\xi (t,t') &\equiv k \int_{t'}^t g(t,s)\Psi(s-t')\, ds\, + \,\th
\int_{t'}^t h(t,s) \Psi(s-t') \,ds\, ,&(A.13c)\cr
}
$$

As an example, let us now apply this general result to the specific case
of $\ddot x = R(t)$. As noted in Eq.~(3.15),
$$
\eqalignno{
u(t)&=x_0+v_0 t \, ,\qquad w(t) = v_0\, ,  &(A.14)\cr
\gtt &= t-t' \, , \qquad\, \htt = 1   \qquad\quad  (t'<t)\, . \cr}
$$
Let us now take $R(t)$ to be an O-U noise, defined by [12, 9]:
$$
\eqalignno{
\po&=\sqrt{\tau \over \pi f_0^2}\exp\left(-{\tau \over f_0^2} y^2\right)
\, .&(A.15a)\cr
T_{\dt}(y_{i+1}\vert y_i) &= \sqrt{\tau \over \pi f_0^2
\left[1-\k^2(\dt)\right]}\exp\left\{- {\tau\left[ (\yy)\right]^2 \over
f_0^2\left[1-\k^2 (\dt)\right]} \right\} \, .&(A.15b)\cr
\k(\dt)&=\exp\left(-{\dt \over \tau}\right)\, .&(A.15c)\cr
}
$$
where $f_0$ is a parameter describing the intensity of the random force.

The Fourier transforms $\widehat P_0$ and $\widehat T$, are (see
Eqs.~(3.4) and (3.9))
$$
\eqalignno{
\widehat P_0 [\zeta]& = \exp\left(-{f_0^2 \over 4\tau} \zeta^2\right)
\, .&(A.16a)\cr
\widehat T_{\dt}(\xi) &= \exp\left[{(1-\k^2)f_0^2 \over
4\tau }\xi^2\right]\, .&(A.16b)\cr
}
$$
{}From Eq.~(A.2), we find
$$
a = - {1 \over \tau}\, ,\eqno(A.17)
$$
so that $\Psi(z)$ is now (from Eq.~(A.5))
$$
\Psi(z) = \exp\left(-{z \over \tau}\right)\, .\eqno(A.18)
$$
{}From the definition of $B(\xi)$, Eq.~(A.8), we have:
$$
B(\xi) = -{f_0^2 \over 2\tau} \xi^2\, .\eqno(A.19)
$$
After some algebra, we obtain
$$
\eqalignno{
\int_{t'}^t h(t,s)\Psi(s-t')\, ds &= \int_{t'}^t
\exp\left[ -{(s-t') \over\tau}\right]\, ds
 = \tau\left\{ 1- \exp\left[ -{(t-t') \over
\tau}\right]\right\}\! ,  &(A.20a)\cr
\int_{t'}^t  g(t,s)\Psi(s-t')\, ds &= \int_{t'}^t (t-s)\exp\left[
-{(s-t') \over \tau}\right]\,ds\cr
&\qquad\quad = \tau\left[t-t'+\tau\left(\exp\left[-{\left(t-t'\right)
\over \tau}\right] -1\right) \right]\, ,&(A.20b)\cr
}
$$
from which we have immediately (by setting $t' = 0$)
$$
\eqalignno{
\int_0^t  h(t,s)\Psi(s)\, ds &= \tau\left[1- \exp\left(-{t \over
\tau}\right)\right] \, , &(A.21a)\cr
\int_0^t  g(t,s)\Psi(s)\, ds &= \tau\left\{t-\tau\left[1-\exp\left(-{t
\over \tau}\right)\right]\right\}\, .&(A.21b)\cr
}
$$
Since $\zeta = k\int_0^t g(t,s)\Psi(s)\, ds + \th\int_0^t
h(t,s)\Psi(s)\,ds$, by substituting (A.21) into (A.16a) we obtain
$$
\eqalignno{
\hat P_0 (\zeta)
&= \exp\left\{-\h \left[a(t)k^2 + b(t)\th^2 +2c(t)k\th
\right]\right\} \, ,&(A.22a)\cr
\noalign{\hbox{where}}
a(t)&={f_0^2 \tau \over 2}\left\{t-\tau\left[1-\exp\left(-{t \over
\tau}\right)\right]\right\}^2 \, , &(A.22b)\cr
b(t) &=  {f_0^2 \tau \over 2}\left[1- \exp\left(-{t \over
\tau}\right)\right]^2 \, , &(A.22c)\cr
c(t)&={f_0^2 \tau \over 2}\left\{ t\left[1- \exp\left(-{t \over
\tau}\right)\right]-\tau\left[ 1-\exp\left(-{t
\over\tau}\right)\right]^2\right\} \, .&(A.22d)\cr
}
$$
Similarly, we must substitute Eqs.~(A.19) and (A.20) into the
expression $ \int_0^t dt' B\left[\xi(t,t')\right]$, where
$\xi(t,t')=k\int_{t'}^t g(t,s)\Psi(s-t')\, ds~+~\th\int_{t'}^t
h(t,s)\Psi(s-t')\, ds\,$. The integral over $t'$ can be performed
explicitly, and after some more algebra , we have that
$$
\eqalignno{
\int_0^t dt' B\left[\xi(t,t')\right] &= -\h \left[l(t)k^2 + m(t)\th^2
+2n(t)k\th \right] \, ,&(A.23a)\cr
\noalign{\hbox{with}}
l(t)&=f_0^2 \left[{t^3 \over 3} + t \tau^2 -t^2\tau +{\tau^3 \over 2} -
2t\tau^2 e^{-t/\tau} -{\tau^3 \over 2}e^{-2t/\tau}\right]\,
.&(A.23b)\cr
m(t)&=f_0^2\left[t-{3 \over 2}\tau +2\tau e^{-t/\tau}-{\tau \over
2}e^{-2t/\tau} \right]\, .&(A.23c)\cr
n(t)&=f_0^2\left[{t^2 \over 2} -t\tau +{\tau^2 \over 2} -\tau^2
e^{-t/\tau} +{\tau^2} e^{-2t/\tau} + t\tau e^{-t/\tau}\right]\,
.&(A.23d)\cr
}
$$

Substituting (A.14), (A.22) and (A.23) into (A.13) yields the
final answer:
$$
\eqalignno{
\hat p(k ,\th ,t)& = \exp\left\{i\left[k(x_0+v_0t) + \th
v_0\right]-\h\left[\alpha(t)k^2 +2\beta(t)k\th + \gamma(t)\th^2
\right]\right\}\,, &(A.24a)\cr
\noalign{\hbox{where}}
\alpha(t)&=f_0^2\left[{t^3 \over 3} -\h t^2\tau
+\tau^3\left(1-e^{-t/\tau}\right) -\tau^2te^{-t/\tau}\right]\, .
&(A.24b)\cr
\beta(t)&=\h f_0^2\left[t^2-\tau
t\left(1-e^{-t/\tau}\right)\right]\, .&(A.24c)\cr
\gamma(t)&=f_0^2\left[t-\tau\left(1-e^{-t/\tau}\right)\right]\,
.&(A.24d)\cr
}
$$

\centerline {\bf Appendix B}\smallskip

To calculate the discrete form of the time derivative of $\fpt$, we
first expand $\hat p(k,\th,n+1)$ in powers of $\dt$, keeping only
terms up to the first order. From the expression for $\fpn$, Eq.~(4.2),
we have
$$
\vbox{\eqalignno{
\hat p(k,\th,n+1) = \exp\Biggl\{ i( & k u_{n+1} +\th w_{n+1})-
\h\biggl[k^2 \sum_{i,j=0}^{n+1} g(n+1,i)g(n+1,j)\p(i,j)\,\,\dt^2\cr
 + &k\th \sum_{i,j=0}^{n+1}
\left[g(n+1,i)h(n+1,j)+ h(n+1,i)g(n+1,j)\right]\p(i,j)\,\,\dt^2\cr
 + &\th^2\sum_{i,j=0}^{n+1} h(n+1,i)h(n+1,j)\p(i,j)\,\,\dt^2
\biggr]\Biggr\}\, .&(B.1)\cr
}}
$$
Let us define $\di g$ as the partial derivative of $g$ with respect to
its first variable, \ie,
$$
\di {g(t,t')} \equiv \pd {g(t,t')} t \, ,\eqno(B.2)
$$
so that $\di {g(n,i)}$ means
$$
\di {g(n,i)} \equiv \pd {g(t,t')} t \big\vert_ {{\scriptstyle {t=n\dt}
\atop \scriptstyle {t'=i\dt}}} \, .\eqno(B.3)
$$
A similar convention holds for the function $h(t,t')$. With this
convention, and for the purpose of calculating the r.h.s of
Eq.~(5.1), we can approximate:
$$
g(n+1,i)g(n+1,j) \approx \left[g(n,i) + \di {g(n,i)} \dt\right]
\left[g(n,j) + \di {g(n,j)} \dt\right] \, .\eqno(B.4)
$$
Keeping terms only up to first order in $\dt$, we have that
$$
\eqalignno{
&\quad g(n+1,i)g(n+1,j)\approx g(n,i)g(n,j) + g(n,j)\di {g(n,i)}\dt + g(n,i)
\di {g(n,j)} \dt \, .&(B.5a)\cr
\noalign{\hbox{}}
& g(n+1,i)h(n+1,j)+h(n+1,i)g(n+1,j)\approx g(n,i)h(n,j) +
h(n,i)g(n,j)\cr
&\qquad + h(n,j)\di {g(n,i)} \dt + h(n,i)\di {g(n,j)} \dt+g(n,j)\di
{h(n,i)} \dt + g(n,i)\di {h(n,j)} \dt \, .\cr
&{}&(B.5b)\cr
\noalign{\hbox{}}
&\quad h(n+1,i)h(n+1,j)\approx h(n,i)h(n,j) + h(n,j)\di {h(n,i)}\dt +
h(n,i)\di {h(n,j)} \dt \, .&(B.5c)\cr
\noalign{\hbox{}}
&\qquad\qquad\qquad\qquad\qquad u_{n+1} \approx u_n + \td {u_n}
t \dt\, .&(B.5d)\cr
\noalign{\hbox{}}
&\qquad\qquad\qquad\qquad\qquad w_{n+1} \approx w_n + \td {w_n}
t \dt\, .&(B.5e)\cr
}
$$
Hence, \eg,
$$
\vbox{\eqalignno{
\sum_{i,j=0}^{n+1} g(n+1,i)h(n+1,j)\p (i,j) \,\dt^2 = &\sum_{i,j=0}^n
g(n,i)h(n,j)\p (i,j) \,\dt^2 \cr
+& \sum_{i,j=0}^n \left[h(n,j)\di {g(n,i)} + g(n,i)\di
{h(n,j)} \right]\p (i,j) \,\dt^3\cr
+&\sum_{j=0}^n g(n+1,n)h(n,j)\p(n,j)\,\dt^2\cr
+&  \sum_{i=0}^n g(n,i)h(n+1,n)\p (i,n)\,\dt^2 \, .&(B.6)\cr
}}
$$
Note that the last term in the l.h.s sum,
$g(n+1,n+1)h(n+1,n+1)\p(n+1,n+1)\,\dt^2$, is of second order in
$\dt$ and was therefore neglected.

Using Eqs.~(B.5), we have that to first order in $\dt$
$$
\eqalignno{
\Delta \hat p(& k,\th,n) = \fpn \cr
&\qquad\qquad \times \left\{ -1 + \exp\left[ i\left(k \td {u(n)} t +
\th \td {w(n)} t\right) -\h \left( Ak^2 +2Bk\th  +
C\th^2\right)\right]\right\}\, ,\cr
&{} &(B.7a) \cr
\noalign{\hbox{where}}
A =& \sum_{i,j=0}^n 2\di {g(n,i)} g(n,j)\p(i,j)\,\dt^3
+2\di g(n,n) \sum_{j=0}^n g(n,j)\p(n,j)\,\dt^2 \, , &(B.7b)\cr
B =& \h \sum_{i,j=0}^n \left[\di {g(n,i)} h(n,j) + \di
{h(n,j)} g(n,i)+ \di {g(n,j)} h(n,i) + \di {h(n,i)} g(n,j)
\right]\p(i,j)\,\dt^3\cr\noalign{\nobreak}
& +\h \di {g(n,n)}\left[ \sum_{i=0}^n
h(n,i)\p(i,n)\,\dt^2 + \sum_{j=0}^n h(n,j)\p(n,j)\,\dt^2\right]\cr
\noalign{\nobreak}
& +\h \di {h(n,n)} \left[\sum_{j=0}^n
g(n,i)\p(i,n)\,\dt^2 +\sum_{i=0}^n g(n,j)\p(n,j)\,\dt^2\right]\, ,
&(B.7c)\cr
C =&  \h \sum_{i,j=0}^n 2\di {h(n,i)} h(n,j)\p(i,j)\,\dt^3
+2\di h(n,n) \sum_{j=0}^n h(n,j)\p(n,j)\,\dt^2\, .&(B.7d)\cr
}
$$

In the limit $\dt \to 0$, we can expand the exponents and keep only the
leading terms. Thus:
$$
\eqalignno{
\Delta &\fpn = \fpn \left[ i\left(k \td {u(n)} t + \th \td {w(n)}
t\right) -\h \left( Ak^2 + 2Bk\th + C \th^2\right)\right]\, ,\cr
&{} &(B.8)\cr
}
$$
where $A, B, C$ are defined in Eqs.~(B.7b)-(B.7d).

In the limit $\li$, all sums become integrals. Taking for example the
coefficient of $k^2$, we have:
$$
\vbox{\eqalignno{
&\left[\sum_{i,j=0}^n 2\di {g(n,i)} g(n,j)\p(i,j)\,\dt^3 +2\di {g(n,n)}
\sum_{j=0}^n g(n,j)\p(n,j)\,\dt^2\right]\cr
&\quad \longrightarrow \left\{\int_0^t \!\!dt'\!\!\int_0^t\!\!
dt''\, \left[2 \di {g(t,t')} g(t,t'') + 2 \di {g(t,t'')} g(t,t')\right]
+ \int_0^t dt' \>\di {g(t,t)} g(t,t')\p(t,t')\right\}\,\dt\, .\cr
 & {}&(B.9a)\cr
}}
$$
The expression on the r.h.s turns out to be
$$
\left\{\td {} t \left[ \int_0^t dt'\int_0^t dt''
g(t,t')g(t,t'')\p(t',t'')\right]\right\}\,\dt \, , \eqno(B.9b)
$$
which is just $d\alpha(t) /dt$, where $\alpha(t)$ is
$$
\alpha (t)= \int_0^t \!\!dt'\!\! \int_0^t\!\! dt''\, g(t,t')g(t,t'')
\,\p(t',t'')\, ,\eqno(B.9c)
$$
This is identical to the definition of $\alpha(t)$ in Eq.~(4.3b).

Similar reasoning applied to the other terms in Eq.~(B.8) leads us to
finally rewrite it, in the limit $\li$, as
$$
\pd {\fpt} t =  \left\{i \td u t k +i\td w t \th -\h \left[
\td \alpha t k^2 +  2 \td \beta t k\th + \td \gamma t \th^2
\right]\right\}\fpt\, ,\eqno(B.10)
$$
where $\alpha, \beta$ and $\gamma$ are
$$
\eqalignno{
\alpha (t)&= \int_0^t \!\!dt'\!\! \int_0^t\!\! dt''\, g(t,t')g(t,t'')
\,\p(t',t'')\, ,&(B.11a)\cr
\beta (t)&= \h \int_0^t\!\! dt'\!\! \int_0^t\!\! dt''\,
\left[g(t,t')h(t,t'')+ h(t,t')g(t,t'')\right]\, \p (t',t'') \, ,
&(B.11b)\cr
\gamma (t)&= \int_0^t\!\! dt' \!\!\int_0^t \!\!dt''\, h(t,t')h(t,t'')
\,\p (t',t'')\, , &(B.11c)\cr
}
$$

\centerline{\bf Appendix C}\smallskip

The function $\fpn$ is given by
$$
\fpn = e^{i(u_nk + w_n\th)}\wn \big\vert_{\f_i = [g(n,i)k +
h(n,i)\th]\dt}\, .\eqno(C.1)
$$
For the Markovian dichotomous noise, we have
$$
\eqalignno{
\po &= \h \left[\delta(y_1-a) + \delta (y_1+a)\right] \, ,&(C.2a)\cr
T_{\dt}(y \vert y') &= \h \left[f(\dt)\delta (y-y') +g(\dt)\delta
(y+y')\right]\, ,&(C.2b)\cr
\noalign{\hbox{where}}
f(\dt) &= 1+ \exp\left(-2\lambda\dt\right)\, ,&(C.2c)\cr
g(\dt) &= 1- \exp\left(-2\lambda\dt\right)\, .&(C.2d)\cr
\noalign{\hbox{and}}
\Wn &= \po \,T_{\dt}(y_2 \vert y_1) \cdots T_{\dt}(y_n \vert
y_{n-1})\, .&(C.2e) \cr
}
$$
We must calculate the Fourier transform of this expression and
substitute the appropriate expressions for the Fourier parameters
$\f_i$. We now define the following two functions
$$
\eqalignno{
W^+_n (y_1,\ldots,y_n)& = \h \delta (y_1-a)\,T_{\dt}(y_2 \vert y_1)
\cdots T_{\dt}(y_n \vert y_{n-1}) \, .&(C.3a)\cr
W^-_n (y_1,\ldots,y_n) &= \h \delta \,(y_1+a)T_{\dt}(y_2 \vert y_1)
\cdots T_{\dt}(y_n\vert y_{n-1}) \, .&(C.3b)\cr
}
$$
According to the definitions of $\po$ and $W_n$ in Eq.~(C.2), we see
that
$$
W_n=W_n^+ + W_n^-\, .\eqno(C.4)
$$
Going over to the Fourier transform of the various functions, we have,
using Eq.~(C.2b),
$$
\vbox{\eqalignno{
\widehat W^+_n (\f_1,\ldots,\f_n) = \Int\!\! dy_1\cdots \Int\!\!
dy_n \,&\exp\left[i(y_1\f_1 + \cdots +y_n\f_n)\right]\cr
&\times \h \delta (y_1-a) \left[\h f\, \delta
(y_2-y_1) + \h g \,\delta (y_2+y_1)\right] \cr
&\times T_{\dt}(y_3 \vert y_2)\cdots T_{\dt}(y_n \vert y_{n-1})\,
.&(C.5)\cr
}}
$$
And a similar equation for $\widehat W^-_n$. After performing the
integration on $y_1$, we have:
$$
\eqalignno{
\widehat W^+_n (\f_1,\ldots,\f_n)= \h e^{ia \f_1}\Int\!\! dy_2\cdots
\Int\!\! dy_n \,&\exp\left[i(y_2\f_2 + \cdots +y_n\f_n)\right]\cr
&\times \left[\h f \delta (y_2-a) + \h g
\delta(y_2+a)\right]\cr
&\times T_{\dt}(y_3\vert y_2)  \cdots T_{\dt}(y_n \vert y_{n-1})\, .
&(C.6a)\cr
\noalign{\hbox{}}
\widehat W^-_n (\f_1,\ldots,\f_n) = \h e^{-ia \f_1}\Int\!\! dy_2\cdots
\Int\!\! dy_n \,& \exp\left[i(y_2\f_2 + \cdots +y_n\f_n)\right]\cr
&\times \left[\h f\, \delta (y_2+a) + \h g \,\delta
(y_2-a)\right]\cr
&\times T_{\dt}(y_3\vert y_2) \cdots T_{\dt}(y_n \vert y_{n-1})\,
.&(C.6b)\cr
}
$$
Using the definitions of $W^+$ and $W^-$, Eq.~(C.3), we have:
$$
\eqalignno{
\widehat W^+_n (\f_1,\ldots,\f_n) &= \h\, e^{ia \f_1} \left\{ f\,
\widehat W^+_{n-1} (\f_2,\ldots,\f_n)+ g \, \widehat W^-_{n-1}
(\f_2,\ldots,\f_n)\right\} \, .&(C.7a)\cr
\widehat W^-_n (\f_1,\ldots,\f_n) &= \h \,e^{-ia \f_1} \left\{ f\,
\widehat W^-_{n-1} (\f_2,\ldots,\f_n)+  g\, \widehat W^+_{n-1}
(\f_2,\ldots,\f_n) \right\}\, .&(C.7b)\cr
\noalign{\hbox{with}}
\widehat W^+_1 &= \h\, e^{i a \f_1}&(C.7c)\cr
\widehat W^-_1 &= \h \,e^{-i a \f_1}&(C.7d)\cr
}
$$
We now have a

\bigskip

{\bf Lemma} : $\widehat W^+_n (\f_1,\ldots,\f_n)=\widehat W^-_n
(-\f_1,\ldots,-\f_n)$.
\bigskip
The proof follows immediately from (C.7) by induction on $n$.

\bigskip

The first terms in the series $\left\{\widehat
W_n^+\right\}_{n=1}^{\infty}$ are:
$$
\eqalignno{
\widehat W^+_1 =& \h e^{i a \f_1} \, .&(C.8a)\cr
\widehat W^+_2 =& \left(\h\right)^2 \left[ f\>e^{i
a\left(\f_1+\f_2\right)} +  g\>e^{i
a\left(\f_1-\f_2\right)}\right]\, .&(C.8b) \cr
\widehat W^+_3 =& \left(\h\right)^3 \Biggl[ f^2\>e^{i
a\left(\f_1+\f_2 +\f_3\right)} + f g\>e^{i
a\left(\f_1+\f_2-\f_3\right)}\cr
&\quad\qquad +f g\>e^{i a\left(\f_1-\f_2-\f_3\right)} +  g^2\>e^{i
a\left(\f_1-\f_2 + \f_3\right)}\Biggr]\, .&(C.8c)\cr
\vdots \quad&\cr
}
$$
This suggests that we can write:
$$
\widehat W^+_n (\f_1,\ldots,\f_n)= \sum_{k=1}^n \left(\h\right)^n
f^{n-k-1}g^{k-1}F^n_k (\f_1, \ldots, \f_n)\, ,\eqno(C.9)
$$
with $F^n_k$ yet to be determined. Note that $\widehat W^-_n
(\f_1,\ldots,\f_n)$ follows from $\widehat W^+_n (\f_1,\ldots,\f_n)$ by
virtue of the above lemma.

Substituting the form (C.9) into the Eqs.~(C.7) we find:
$$
\eqalignno{
\sum_{k=1}^n \left(\h\right)^n \!\!\!\!f^{n-k-1}g^{k-1}F^n_k(\f_1,
\ldots, \f_n)= &\h f e^{i a \f_1}\sum_{j=1}^{n-1} \left(\h\right)^{n-1}
f^{n-j-2}g^{j-1}F^{n-1}_j(\f_2, \ldots, \f_n)\cr
 + &\h g e^{ia\f_1}\sum_{j=1}^{n-1} \left(\h\right)^{n-1}\!\!\!\!
f^{n-j-2}g^{j-1}F^{n-1}_j(-\f_2, \ldots, -\f_n)\, .\cr
&{}&(C.10) \cr
}
$$
The first term on the r.h.s. can be rewritten as
$$
e^{i a\f_1}\sum_{k=1}^{n-1}\left(\h\right)^{n}f^{n-k-1}g^{k-1}
F^{n-1}_k(\f_2,\ldots, \f_n) \, , \eqno{(C.11a)}
$$
where the index $j$ has been renamed $k$. The second term on the r.h.s
of Eq.~(C.10) can be rewritten as
$$
e^{ia\f_1}\sum_{k=2}^{n}\left(\h\right)^{n}f^{n-k-1}g^{k-1}
F^{n-1}_{k-1} (-\f_2,\ldots, -\f_n) \, , \eqno{(C.11b)}
$$
where the new summation index $k$ is defined as $k=j+1$.

Comparing the two sides of Eq.~(C.10) term by term with the help of
Eqs.~(C.11), we see that we must have:
$$
F^{n}_k(\f_1,\ldots, \f_n)
=\cases{e^{ia\f_1}F^{n-1}_1(\f_2,\ldots,\f_n)&$k=1$\cr
{}&$\quad$\cr
e^{ia\f_1}\left[F^{n-1}_k(\f_2,\ldots,\f_n)
+F^{n-1}_{k-1}(-\f_2,\ldots,-\f_n)\right]&$1<k<n$\cr
{}&$\quad$\cr
e^{ia\f_1}F^{n-1}_{n-1}(-\f_2,\ldots,-\f_n)&$k=n\, .$\cr}
\eqno(C.12)
$$
The three cases in Eq.~(C.12) can be summed up as
$$
\eqalignno{
F^n_k(\f_1,\ldots, \f_n)& = e^{ia\f_1}\left[F^{n-1}_k(\f_2,\ldots,\f_n)
+F^{n-1}_{k-1}(-\f_2,\ldots,-\f_n)\right] \, ,&(C.13a)\cr
\noalign{\hbox{if we add the conventions}}
F^m_0(\f_1,\ldots, \f_n)&=0\, .&(C.13b)\cr
F^m_{m+1}(\f_1,\ldots, \f_n)&=0\, .&(C.13c)\cr
}
$$
Finally, we note that from $\widehat W^+_1 = \h e^{i a \f_1}$, we have
that
$$
F^1_1(\f_1)=e^{ia\f_1}\, .\eqno(C.14)
$$

Eqs.~(C.13) and (C.14) fully determine the coefficients
$F^n_k(\f_1,\ldots, \f_n)$. Eqs.~(C.1), (C.4), (C.9) and the lemma now
allow us to write the function $\fpn$ as
$$
\vbox{\eqalignno{
\fpn &= \exp\left[i(k u_n +\th w_n)\right]
\sum_{k=1}^n \left(\h\right)^n f^{n-k-1}g^{k-1}\cr
&\times  \left[F^n_k(\f_1,
\ldots, \f_n) + F^n_k(-\f_1, \ldots, -\f_n)\right]
\big\vert_{\f_i=\left[k g(n,i)+\th h(n,i)\right]\dt} \, ,&(C.15)\cr
} }
$$
\centerline{\bf Appendix D}\smallskip

We wish to calculate the discrete version of the first derivative,
$\Delta\fpn$, defined as $\widehat p(k,\th,n~+~1)~-~\fpn$, with
$\fpn$ defined by Eq.~(C.15). Rewriting (C.15) with $n$ replaced by
$n+1$ yields
$$
\vbox{\eqalignno{
\hat p(k,\th,n+1) = &\sum_{k=1}^{n+1}\Biggl\{\left(\h\right)^{n+1}
f^{n-k}g^{k-1} \cr
&\times\left[F^{n+1}_k(\f_1, \ldots, \f_{n+1}) + F^{n+1}_k(-\f_1,
\ldots, -\f_{n+1})\right]\biggr\}\cr
=&\sum_{k=1}^{n+1}\left(\h\right)^{n+1}\!\!\! f^{n-k}g^{k-1}\cr
&\times \biggl\{e^{ia\f_1}[F^n_k(\f_2, \ldots, \f_{n+1}) + F^n_k(-\f_2,
\ldots, -\f_{n+1})]\cr
& + {e^{-ia\f_1}[F^n_k(-\f_2, \ldots,-\f_{n+1}) + F^n_k(\f_2,
\ldots, \f_{n+1})]}\Biggr\}\, , &(D.1)\cr
}}
$$
where we have used the recurrence relations for $F^n_k$, Eqs.~(C.13).
In Eq.~(D.1), $ \f_i=[~k~g~(~n~+~1~-~i~)+\th~h~(~n~+~1~-~i~)~]~\dt$,
in accordance with our assumption that $g(n,i)=g(n-i)$.

Remembering that $F_{n+1}^n(\f_2, \ldots, \f_{n+1})=0$ (see
Eq.~(C.13c)), we can rewrite the first and third terms in the r.h.s of
Eq.~(D.1) as:
$$
\vbox{\eqalignno{
\sum_{k=1}^{n+1}&\left(\h\right)^{n+1}\!\!\!
f^{n-k}g^{k-1}\left[e^{ia\f_1} F^n_k(\f_2, \ldots, \f_{n+1}) +
e^{-ia\f_1}F^n_k(-\f_2, \ldots, -\f_{n+1})\right]\cr
&=\sum_{k=1}^n \left(\h\right)^n f^{n-k-1}g^{k-1} {f \over 2}
\left[e^{ia\f_1}F^n_k(\f_2, \ldots, \f_{n+1}) + e^{-ia \f_1}F^n_k(-\f_2,
\ldots, -\f_{n+1})\right]\, .\cr
&{}&(D.2)\cr
}}
$$
In the remaining two terms in the r.h.s of Eq.~(D.1), we change the
summation index from $k$ to $j=k-1$. Remembering that $F_0^n(\f_2,
\ldots, \f_{n+1}) =0$ (see Eq.~(C.13b)), we have:
$$
\vbox{\eqalignno{
\sum_{k=1}^{n+1}&\left(\h\right)^{n+1}\!\!\!
f^{n-k}g^{k-1}\left[e^{ia\f_1}F^n_k(-\f_2, \ldots, -\f_{n+1}) +
e^{-ia\f_1}F^n_k(\f_2, \ldots, \f_{n+1})\right]\cr
&=\sum_{j=1}^n \left(\h\right)^n f^{n-j-1}g^{j-1} {g \over 2}
\left[e^{-ia\f_1}F^n_j(\f_2, \ldots, \f_{n+1}) + e^{ia \f_1}F^n_j(-\f_2,
\ldots, -\f_{n+1})\right]\, .\cr
&{} &(D.3)\cr
}}
$$
Combining (D.2) and (D.3), we obtain:
$$
\vbox{\eqalignno{
\hat p(k,\th,n+1)= \sum_{k=1}^n &\left(\h\right)^n
f^{n-k-1}g^{k-1}\cr
&\times \biggl\{{f \over 2}\left[e^{ia\f_1}F^n_k(\f_2, \ldots,
\f_{n+1}) + e^{-ia\f_1}F^n_k(-\f_2, \ldots, -\f_{n+1})\right]\cr
&+ {g \over 2}\left[e^{-ia\f_1}F^n_k(\f_2, \ldots, \f_{n+1}) +e^{ia \f_1}
F^n_k(-\f_2,  \ldots, -\f_{n+1})\right]\biggr\}\, ,\cr
{}& &(D.4)\cr
}}
$$
with $\f_i= k g(n-i+1) + \th h(n+1-i)$. From this we subtract
$$
\fpn=\sum_{k=1}^n\left(\h\right)^n f^{n-k-1}g^{k-1}
\left[F^n_k(\varphi'_1, \ldots, \varphi'_n) + F^n_k(-\varphi'_1, \ldots,
-\varphi'_n)\right] \, ,\eqno(D.5)
$$
where $\varphi'_i= k g(n-i) + \th h(n-i) $. Note now that
$\f_{i+1}=\varphi'_i $ (it is for this relation that we assumed
that $g(n,i)=g(n-i)$ and $h(n,i)=h(n-i)$). Hence, $\Delta \hat p $ can
be written as:
$$
\vbox{\eqalignno{
\Delta \hat p = \sum_{k=1}^n\left(\h\right)^n f^{n-k-1}g^{k-1}
\Biggl\{& F^n_k(\f_2, \ldots, \f_{n+1})\left[{f \over 2} e^{ia\f_1} + {g
\over 2} e^{-ia\f_1} - 1\right]\cr
+ & F^n_k(-\f_2, \ldots, -\f_n)\left[{f\over 2} e^{-ia\f_1} + {g \over 2}
e^{ia\f_1} -1 \right]\Biggr\}\, .&(D.6)\cr
} }
$$
Since ultimately we want the limit $\dt \to 0$ of $\Delta \hat p /
\dt$, we now expand the expressions in square brackets up to first order
in $\dt$. Referring to Eq.~(C.2), we have
$$
\eqalignno{
f(\dt) &= 2 -2\lambda \dt + 2\lambda^2\dt^2 + \ldots &(D.7a)\cr
g(\dt) &= 2\lambda \dt - 2\lambda^2\dt^2 + \ldots&(D.7b)\cr
\exp(ia\f_1)& = 1+ ia\f_1 - \h a^2 \f^2_1 + \ldots&(D.7c)\cr
}
$$
where the last line follows from the fact that $\f_1$ is first order in
$\dt$, because $\f_1=\left[k g(n) + \th h(n)\right]\dt $.

Hence, to first order in $\dt$:
$$
\Delta \fpn = ia\f_1 \sum_{k=1}^n\left(\h\right)^n\!\!
f^{n-k-1}g^{k-1} \left[F^n_k(\f_2, \ldots, \f_{n+1}) - F^n_k(-\f_2,
\ldots, -\f_{n+1})\right] \, .\eqno(D.8)
$$
This expression is not related in any obvious way to $\fpn$ and therefore
we cannot write a first order differential equation for $\fpn$. Let us
therefore look at the second derivative of $\fpn$, the discrete form of
which is $\Delta^2 \hat p / \dt^2 $ where $\Delta^2 \hat p$ is given by
$\hat p(k,\th,n+2) + \fpn -2\hat p(k,\th,n+1)$. The expression
$\hat p(k,\th,n+2)$ is calculated in much the same way as $\hat
p(k,\th,n+1)$ in Eqs.~(D.1)-(D.4). This time, we'll need to use the
recurrence relation Eq.~(C.13) twice. The algebra is straightforward,
and we obtain finally:
$$
\vbox{\eqalignno{
\hat p & (k,\th,n+2)=\sum_{k=1}^n \left(\h\right)^n
f^{n-k-1}g^{k-1}\cr
&\times \Biggl\{F^n_k(\f_2, \ldots, \f_{n+1})\biggl[{f^2 \over 4}
e^{ia(\f_0+\f_1)} + {fg \over 4} e^{ia(\f_1-\f_0)}
 + {fg \over 4}e^{-ia(\f_1+\f_0)} + {g^2 \over 4} e^{ia(\f_0
-\f_1)}\biggr]\cr
& + F^n_k(-\f_2, \ldots, -\f_n)\left[{f^2 \over 4}
e^{-ia(\f_0+\f_1)} + {fg \over 4} e^{ia(\f_0-\f_1)} + {fg \over 4}
e^{ia(\f_0+\f_1)} + {g^2 \over 4} e^{ia(\f_1 -\f_0)}\right]\Biggr\}\, ,\cr
{}& &(D.9)\cr
}}
$$
where, for the sake of consistency, $\f_1,\ldots,\f_{n+1}$ are the {\it
same} as in Eq.~(D.1), \ie, $\f_i=[k g(n-i+1) + \th h(n-i+1)]\dt $,
and we have defined $\f_0=\left[k g(n+1) + \th h(n+1)\right]\dt $.

Substituting Eqs.~(D.4), (D.5) and (D.9) into the expression for
$\Delta^2 \hat p$ and expanding to second order in $\dt$ (using
Eq.~(D.7)), we finally obtain:
$$
\vbox{\eqalignno{
\Delta^2 \fpn = &ia\left[(\f_0 - \f_1) -2\lambda \f_0
\dt\right]\cr
&\times\left\{\sum_{k=1}^n\left(\h\right)^n
f^{n-k-1}g^{k-1} \left[F^n_k(\f_2, \ldots, \f_{n+1}) - F^n_k(-\f_2,
\ldots, -\f_{n+1})\right]\right\}\cr
&- a^2\f_1^2  \sum_{k=1}^n\left(\h\right)^n
f^{n-k-1}g^{k-1} \left[F^n_k(\f_2, \ldots, \f_{n+1}) + F^n_k(-\f_2,
\ldots, -\f_{n+1})\right]\, .\cr
{}& &(D.10)\cr
}}
$$
Note that $\f_0 - \f_1$ is indeed of second order in $\dt$, as it should
be, because $\f_0 - \f_1 = \left\{k\left[g(n+1) - g(n)\right] +
\th\left[h(n+1) - h(n)\right]\right\}~\dt$, and $g(n+1)-g(n)$ is
already of first order in $\dt$. Comparing Eq.~(D.10) with Eqs.~(D.5)
and (D.8), we have that
$$
{\Delta^2 \fpn \over \dt^2} = -{a^2\f_1^2 \over \dt^2}\, \fpn + \left[
{\f_0 -\f_1 -2\lambda \f_0 \dt \over \f_1 \dt}\right] { \Delta \fpn
\over \dt} \, .\eqno(D.11)
$$
Denoting $\f \equiv k g(t) + \th h(t)$, we see that
$$
\eqalignno{
&\qquad{\f_1^2 \over \dt^2} = \left[k g(n) + \th h(n) \right]^2 \to
\f^2 \, .&(D.12a)\cr
\noalign{\hbox{}}
{\f_0-\f_1 \over \f_1 \dt}
 =& {1 \over \f_1}\left[k  {g(n+1)-g(n) \over \dt} +\th {h(n+1)-h(n)
\over \dt}\right]\to {1 \over \f} \td \f t\, .&(D.12b)\cr
\noalign{\hbox{}}
&\qquad\qquad\quad {\f_0 \dt \over \f_1 \dt} \to 1\, .&(D.12c)\cr
}
$$
where we have used the fact that $\f_0, \f_1 \to \f$ when $\li$. Hence,
in the limit $\li $, Eq.~(D.10) finally becomes:

$$
\eqalignno{
&\pdd {\fpt} t +\left(2\lambda - {1\over \f} \td \f t\right)\pd {\fpt} t
+ a^2\f^2\fpt = 0  \, ,&(D.13a)\cr
\noalign{\hbox{where}}
&\f \equiv k g(t) + \th h(t)\, . &(D.13b)\cr
}
$$

\centerline{\bf References}\smallskip

\refi The classic review of the standard theory is S.~Chandrasekhar,
\rmp 15 1 1943 , reprinted in {\it Selected Papers on noise and
Stochastic Processes}, edited by N.~Wax (Dover, New York, 1964).
\refi M.~I.~Dykman, R.~Mannella, P.~V.~E.~McClintock, S.~M.~Soskin and
N.~G.~Stocks, \pra 43 1701 1991 .
\refi M.~I.~Dykman, \pra 43 2020 1989 ; M.~I.~Dykman, D.~G.~Luchinsky,
P.~V.~E.~McClintock, N.~D.~Stein and N.~G.~Stocks, \pra 46 R1713 1992 .
\refi N.~G.~Van~Kampen, \jsp 54 1289 1989 .
\refi {\it Noise in Nonlinear Dynamical Systems}, edited by F.~Moss and
P.~V.~E.~McClintock (Cambridge University, Cambridge, 1989), Vol I and
references therein. See particularly the contributions of J.~M.~Sancho
and M.~San~Miguel (p. 72) and P.~H\"anggi (p. 307).
\refi See, \eg, B.~J.~Alder and T.~E.~Wainwright, \prl 18 988 1967 ;
\pra 1 18 1970 .
\refi R.~Kubo, M.~Toda and N.~Hashitsume, {\it Statistical Physics II}
(Springer-Verlag, Berlin, 1985).
\refi M.~Araujo, S.~Havlin, H.~Larralde and H.~E.~Stanley, \prl 68 1791
1992 .
\refi J.~Heinrichs, \pre 47 3007 1993 .
\refi J.~Masoliver, \pra 45 706 1992 .
\refi J.~Masoliver, \pre 48 7121 1993 .
\refi N.~G.~Van~Kampen, {\it Stochastic Processes in Physics and
Chemistry}, (2nd edition, North Holland, Amsterdam, 1992).
\refi E.~g., R.~Ghez, {\it A Primer of Diffusion Problems} (Wiley, New
York, 1988) p.187.
\refi H.~Risken, {\it The Fokker-Planck Equation} (2nd edition,
Springer-Verlag, Berlin, 1989).
\refi J.~Heinrichs, \pre 48 2397 1993 .
\refi R.~F.~Pawula, \pra 35 3102 1987 ; \pra 36 4996 1987 .

\vfill\eject\bye

\end